\documentclass[aps,preprint]{revtex4}

\usepackage{amsmath,amssymb}
\usepackage[bookmarks=true]{hyperref}
\def\gfxon{\usepackage[final]{graphicx}}

\gfxon
\usepackage{subfigure}

\usepackage[T1]{fontenc}
\usepackage{amsthm}
\usepackage{amsfonts}
\usepackage{bm}
\usepackage{cancel}
\usepackage{array}

\makeatletter
\let\old@startsection=\@startsection
\renewcommand{\@startsection}[6]{\old@startsection{#1}{#2}{#3}{#4}{#5}{#6\mathversion{bold}}}
\makeatother

\newcommand {\de}{\partial}
\newcommand {\Tr}{{\rm Tr}\,}
\newcommand{\elm}{\textsc{em}}
\makeatletter
\def\mr@ignsp#1 {\ifx\:#1\@empty\else #1\expandafter\mr@ignsp\fi}%
\newcommand{\multiref}[1]{\begingroup
\xdef\mr@no@sparg{\expandafter\mr@ignsp#1 \: }%
\def\mr@comma{}%
\@for\mr@refs:=\mr@no@sparg\do{\mr@comma\def\mr@comma{,}\ref{\mr@refs}}%
\endgroup}
\makeatother

\ifx\href\asklfhas\newcommand{\href}[2]{#2}\fi

\begin{document}

\title{Colorful boojums at the interface of a color superconductor}

\author{ Mattia Cipriani$^{1}$, Walter Vinci$^{1,2}$, Muneto Nitta$^{3}$}

\affiliation{$^1$University of Pisa, Department of Physics ``E. Fermi'', INFN,
Largo Bruno Pontecorvo 7, 56127 Pisa, Italy \\
$^2$ London Centre for Nanotechnology, University College London,\\
17-19 Gordon Street, London, WC1H 0AH,
United Kingdom\\
$^3$Department of Physics, and Research and Education Center 
for Natural Sciences, Keio University, 4-1-1 Hiyoshi, Yokohama, 
Kanagawa 223-8521, Japan
}

\hyphenation{superconducting}

\begin{abstract}
	We study junctions of vortices, or boojums, at the interface between color and hadronic superconducting/superfluid phases. This type of interface could be present in the interior of neutron stars, where an inner core made of quark matter in the color-flavor-locked  phase  is surrounded by an outer shell of superconducting protons and superfluid neutrons. We study the fate of magnetic (proton) and superfluid (neutron) vortices as they enter the color-flavor locked phase. 
We find that proton vortices terminate on Dirac monopoles of the massless magnetic field, and magnetic fluxes of massive gauge field spread along the surface 
and are screened by surface superconducting currents. 
On the other hand, three neutron vortices join 
at a boojum and are split into three color magnetic vortices
with different color magnetic fluxes canceled out in total. The color magnetic vortices  host confined color-magnetic monopoles when strange quark mass is taken into account. We also present a simple numerical model of the shape of the neutron boojum.
\end{abstract}

%
%
%
%
%
%
%
%
%
%
%
%
%
%
%
%
%

\maketitle

\section{Introduction}
Superfluidity is a remarkable quantum phenomenon appearing in 
a wide range of physical systems, from helium superfluid \cite{Volovik:2003} and 
ultra-cold atomic gasses \cite{Pitaevskii:2003}, 
to quantum chromo-dynamics (QCD) and 
cosmology \cite{Vilenkin:2000}. 
One of the outstanding consequences of superfluidity is 
the existence of quantized vortices.  
The observation of quantized vortices gives a direct evidence of superfluidity; 
for instance, ultra-cold atomic Bose-Einsten condensates (BECs) 
were proved to be superfluids 
by the direct observation of an Abrikosov vortex lattice under the rotation 
\cite{Abo-Shaeer:2001}. 
The superfluidity of fermionic atomic gasses 
was also shown in the all range of BEC/BCS crossover 
by observing the vortex lattices \cite{Zwierlein:2005}. 
Superfluid vortices also play central roles in quantum turbulence 
in superfluid helium and atomic BECs \cite{Tsubota:2012}.

When vortices cross an interface between two distinct superfluid phases,
they may form very interesting structures called
``boojums'', which appear around the interface \cite{Mermin}. 
Various types of boojums have been already studied  
in nematic liquids \cite{Volovik:1983}, 
superfluids at the edge of a container filled with $^4$He, 
at the A-B phase boundary of $^3$He \cite{Blaauwgeers:2002}, 
in multi-component or spinor Bose-Einstein condensates 
\cite{Takeuchi:2006,Kasamatsu:2010aq, Borgh:2012}.
Such boojums are also considered in field theories 
such as non-linear sigma models \cite{Gauntlett:2000de} and 
$U(1)$ gauge-theories \cite{Shifman:2002jm}.

In this Letter we report our finding of 
``colorful boojums''  appearing at the interface \cite{Giannakis:2003am,Alford:1999pb,Alford:2001zr} 
of a color superconductor.  
It is most likely that the $npe$ phase exists in the core of neutron stars, 
where neutrons and protons are 
superfluid and superconducting, respectively. 
Since neutron stars are rotating rapidly and are accompanied by large magnetic fields, 
both superfluid and superconducting vortices exist in the $npe$ phase \cite{neutron-star}.
Such vortices are expected to explain the pulsar glitch phenomenon \cite{Anderson:1975zze}. 
On the other hand, it has been shown that, at extremely high density, 
hadrons are converted into quark matter which exhibits color superconductivity as well as superfluidity \cite{Alford:1998mk,Rajagopal:2000wf,Alford:2007xm}.
This phase of QCD, also called  color-flavor-locked phase (CFL), 
may exist in the inner core of neutron stars. 
When the matter in the CFL phase is rotating, such as in the core of neutron stars, 
superfluid vortices \cite{Forbes:2001gj,Iida:2002ev} 
are inevitably created and expected to constitute a vortex lattice. 
However, the simplest superfluid vortex carrying integer circulation is unstable and decay \cite{Nakano:2007dr,Forbes:2001gj,Iida:2002ev} into a set of three non-Abelian vortices 
\cite{Balachandran:2005ev}, each of which carries a color-magnetic flux and 
1/3 quantized circulation \cite{Balachandran:2005ev,Eto:2009kg}.
It is non-Abelian vortices which constitute a lattice \cite{Nakano:2007dr} 
and may play various roles in the inner core of a neutron star \cite{Sedrakian:2008ay}
if the CFL phase is actually present there.

We show that boojum structures are created when  
superfluid and superconducting vortices of the $npe$ phase
penetrate into the CFL core. 
We find that the proton boojums are accompanied by Dirac monopoles and 
surface superconducting currents 
while neutron boojums are accompanied by confined color-magnetic monopoles of two different kinds. We also obtain the shape of the neutron boojum split into three color magnetic vortices.
\section{Vortices in the CFL Phase}

The CFL phase is characterized by the di-quark condensate \cite{Alford:2007xm,Alford:1998mk}:
\begin{align}
	\left<\psi_{i}^{\alpha}C\gamma_{5}\psi_{j}^{\beta}\right> 
=  \epsilon^{\alpha \beta \gamma}\epsilon_{ijk}\Phi_{\gamma}^{\ k} ,
\end{align} 
where $i,j,k=u,d,s$ are flavor indices, while $\alpha,\beta,\gamma=r,g,b$ are color indices. The quarks pair in a parity-even, spin singlet channel, while the color and flavor wave functions are completely antisymmetric. In the ground state of the CFL phase, the order parameter takes the form:
\begin{equation}
	\left<\Phi_{\gamma}^{\ k}\right>= \Delta_{\textsc{cfl}}\delta_{\gamma}^{\ k} \, ,
\end{equation}
up to color-flavor rotations.
If we neglect quark massess, the pattern of symmetry breaking in this ground state is:
\begin{equation}\label{eq:QCDsymmetry}
	U(1)_{B} \times SU(3)_{C} \times SU(3)_{F} \rightarrow SU(3)_{C+F} \, ,
\end{equation}
apart from a discrete symmetry.
The residual color-flavor symmetry is one of the peculiar properties of the CFL phase, which gives raise to a very interesting physics.
The symmetry breaking pattern allows for the existence of stable semi-superfluid non-Abelian vortices \cite{Balachandran:2005ev}, topologically characterized by:
\begin{equation}
	\pi_{1} \left( \frac{U(1)_{B} \times SU(3)_{C}}{\mathbb{Z}_{3}} \right) = \mathbb{Z} \, .
\end{equation} 
These vortices wind both in the $SU(3)_{C}$ and $U(1)_{B}$ groups,  
and consequently  
they carry color magnetic fluxes and quantized circulation.
A peculiarity of these vortices is that they further break the residual color-flavor symmetry 
to its subgroup $SU(2)_{C+F} \times U(1)_{C+F}$. 
This breaking allows the existence of 
Nambu-Goldstone modes 
localized along the vortex core \cite{Nakano:2007dr,Eto:2009bh,Hirono:2010gq}:  
\begin{equation}
\mathbb CP^{2} = \frac{SU(3)}{SU(2)\times U(1)}.
\label{eq:moduli}
\end{equation}
These modes are collective coordinates of the vortex, 
that is,  gapless excitations propagating along the vortex.


It turns out that  flavor symmetry already includes  the electromagnetic group $U(1)_{\elm}$, because its generator is proportional to one of the generators of the flavor group $SU(3)_{F}$:
	$T^{\elm}= \frac13 {\rm diag}(-2,1,1)
\propto T^{8} \in \mathfrak{su}(3)_{F} $.
When electromagnetic interactions are taken into account and the  $U(1)_{\elm}$ is gauged, 
the $SU(3)_{F}$ symmetry is reduced as 
$SU(3)_{F}\stackrel{T^{\elm}}{\longrightarrow }SU(2)_{F}\times U(1)_{\elm}$, 
and the unbroken color-flavor group is $SU(2)_{C+F}$.
Moreover, in the CFL ground state the action of color and flavor rotations is indistinguishable, and the proportionality relation 
above allows for the mixing of the $A^{\elm}$ and $A^{8}$ fields, 
generating massless and massive combinations \cite{Rajagopal:2000wf}:
\begin{eqnarray}\label{eq:mixing}
  & A_{0} = -\sin \zeta A^{\elm}+\cos \zeta A^{8} , 
    A_{\rm M} = \cos \zeta A^{\elm}+\sin \zeta A^{8} , 
	 \nonumber \\
   & \cos\zeta\equiv\sqrt{\frac{e^{2}}{e^{2}+3 g^{2}_{s}/2}}\equiv\frac{e}{g_{\rm M}},
\hspace{-1cm}
\end{eqnarray}
with gauge couplings $e$ and $g$ of $U(1)_{\elm}$ and $SU(3)_C$, respectively. 
%

As we showed in \cite{Vinci:2012mc}, when the electromagnetic interactions are included the degenerate set (\ref{eq:moduli}) is partially lifted by a small potential. There are two types of vortices left, which wind differently inside the color group. The ``Balachandran-Digal-Matsuura (BDM)'' vortex \cite{Balachandran:2005ev}, that  winds only along $T^{8}$  inside $SU(3)_{C}$, and the ``$\mathbb{C}P^{1}$'' vortices, winding also along $T^{3}$.  $\mathbb{C}P^{1}$ vortices are related among themselves by the residual $SU(2)_{C+F}$. We distinguish between the $\mathbb{C}P^{1}_{+}$ vortex, winding along  $+T^{3}$, and the $\mathbb{C}P^{1}_{-}$ vortex winding along $-T^{3}$. We take a gauge in which the BDM, $\mathbb{C}P^1_+$ and $\mathbb{C}P^1_-$  vortices carry color fluxes $\bar{r}$, $\bar{g}$ and $\bar{b}$, respectively.

These vortices are peculiar solutions of the Ginzburg-Landau (GL)  free energy of the CFL phase 
valid for temperatures near the critical one, $T_{c}$ 
\cite{Giannakis:2001wz,Iida:2002ev,Iida:2000ha,Iida:2003cc,Yasui:2010yw}.
Since the interesting vortex configurations  are built out of gauge  fields corresponding to generators commuting with $T^{8}$ \cite{Vinci:2012mc}, we can consistently consider the following reduced form of the GL Lagrangian:
\begin{align}\label{eq:LGlagrangian}
	\mathcal L_{\rm GL} & = \Tr\left[-\frac14\frac32 F^{0}_{ij}F^{0ij} - \frac14\frac32 F^{M}_{ij}F^{Mij} -\frac14 F^{b}_{ij}F^{bij} \right]  \nonumber \\	
	& \quad - \Tr\left[ K_{3}\nabla_{i} \Phi^{\dagger}\nabla^{i} \Phi + m^{2}\Phi^{\dagger}\Phi \right] +V_{\rm GL} \, , \\
	V_{\rm GL} &= -\lambda_{2}\Tr(\Phi^{\dagger}\Phi)^{2} - \lambda_{1}(\Tr[\Phi^{\dagger}\Phi])^{2}-\frac{3 m^{4}}{4(3\lambda_{1}+\lambda_{2})} \, ,\nonumber
\end{align}
written in terms of the massive and massless combinations (\ref{eq:mixing}). 
Here we have defined 
$\nabla_{i} = \de_{i}-ig_{\rm M}A^{\rm M}T^{\rm M}-i g_{s}A^{b}T^{b} $, $T^M=\frac13 {\rm diag}(-2,1,1)$ and 
$g_{\rm M} = \sqrt{e^{2}+3 g_{s}^{2}/2}$, 
where $b$ is an index of $SU(2)_{C}$ satisfying $ [T^{b},T^{8}]=0 $.
Because we are interested only in static solutions, 
we omitted the terms involving time derivatives in Eq. \eqref{eq:LGlagrangian}.
From Eq. \eqref{eq:LGlagrangian} we can see that the massless field $A_{0}$ decouples and does not interact with the vortices in the diagonal entry. 
The BDM solution is described in the polar coordinates $(r,\theta)$ as
\begin{align}
	& \Phi(r,\theta)_{\textsc{bdm}}= \Delta_{\textsc{cfl}}
		{\rm diag}(e^{i\theta}f(r),g(r),g(r))  , \nonumber \\
	& A_{i}^{\rm M}T^{\rm M}= \frac1{g_{\rm M}}\frac{\epsilon_{ij} x^{j}}{r^{2}}[1-h(r)] \, T^{\rm M} , \; A_{i}^{0}=0 \,.
	\label{eq:BDMvortex}
\end{align}
The $\mathbb{C}P^{1}_{+}$ vortex is instead described by
\begin{align}
	\Phi(r,\theta)_{\mathbb CP^{1}_{+}} &= \Delta_{\textsc{cfl}}
			{\rm diag}(g_1(r), e^{i\theta}f(r),g_2(r))  , \nonumber \\
	A_{i}^{\rm M}T^{\rm M} &= - \frac12 \frac1{g_{\rm M}} \frac{\epsilon_{ij} x^{j}}{r^{2}}[1-h(r)] T^{\rm M} ,\nonumber \\
	A_{i}^{3}T^{3} &= \frac{1}{\sqrt2} \frac1{g_{s}} \frac{\epsilon_{ij} x^{j}}{r^{2}}[1-l(r)]\,T^{3}\,,
\end{align}
and the $\mathbb{C}P^{1}_{-}$, having the winding in the remaining entry and the same tension, is obtained by $SU(2)_{C+F}$ transformations. 

The low-energy effective action for  
the gapless excitations in Eq.~(\ref{eq:moduli}) along the vortex line 
(in the $x^3$ direction) can 
be described by a 1+1 dimensional vortex world-sheet theory \cite{Eto:2009bh}
\begin{align}\label{eq:effaction}
	\mathcal{L}_{\mathbb{C}P^{2}} 
 = C \sum_{\alpha=0,3} {K_{\alpha} \left[\de_{\alpha}\phi \de^{\alpha}\phi 
      + (\phi^{\dagger} \de_{\alpha}\phi)(\phi^{\dagger} \de^{\alpha}\phi)\right]} ,  
\end{align}
where we have introduced the homogeneous coordinates $\phi=(\phi_{1},\phi_{2},\phi_{3})^{T}$ 
of $\mathbb{C}P^{2}$, subject to the condition $\phi^{\dagger} \phi~=~1$, 
and we have restored time derivatives for completeness.
Here, $C$ is a finite constant \cite{Eto:2009bh} and $K_{0,3}$ is obtained from weak coupling calculations as $K_0 = 3K_{3} \sim \mu^{2}/T_{c}^{2}$.  
The isometry corresponding to $T^8 \propto T^{\elm}$ is gauged \cite{Hirono:2012ki},
but we do not need it for our study.

The BDM vortex corresponds to $\phi=(1,0,0)$, the $\mathbb{C}P^{1}_{+}$ one to $\phi=(0,1,0)$ and the $\mathbb{C}P^{1}_{-}$ one to $\phi=(0,0,1)$. 
The BDM and $\mathbb{C}P^{1}$ vortices are separated by a potential energy barrier due to the electromagnetic interactions \cite{Vinci:2012mc}. When the mass of the strange quark is negligible with respect to the chemical potential, this potential is the relevant 
one (larger than the quantum mechanically induced one \cite{Eto:2011mk}), 
and the BDM solution proves to have the lowest tension, while $\mathbb{C}P^{1}$ vortices have a slightly higher energy, which makes them metastable. 
However, the effect of a very small but not negligible strange quark mass, in the limit of $\mu \gg m_s^2/\Delta_{\textsc{cfl}}$, can be considered as a perturbation and an additional term has to be included in the Ginzburg-Landau theory
\cite{Iida:2003cc}:
	$V_{\textsc{s}} = \epsilon \Tr \left[ \Phi^{\dagger} \left( \frac23 \mathbb{I}_{3} + T^{3} \right) \Phi \right]$ 
with $\epsilon \propto m_s^2$. 
The implications of this potential for the vortex solutions can be seen by looking at the effective action for pure color vortices, neglecting the electromagnetic coupling.
The effective potential is induced \cite{Eto:2009tr}: 
 \begin{align}
	V_{\mathbb{C}P^{2}}^{\rm eff} &= \epsilon \int d^2 x \Tr \left[ \Phi^{\dagger} T^{3} \Phi \right] \equiv D \left( |\phi_{3}|^{2} - |\phi_{2}|^{2} \right) \, \label{eq:strangepotential} ,
\end{align}
where the coefficient $D$ is given 
by using the vortex profile functions as
	$D= \pi \epsilon \Delta_{\epsilon}^2 \int_{0}^{\infty} dr \, r \, (f^{2}-g^{2})$, 
with $\Delta_{\epsilon} = \Delta_{\textsc{cfl}}(m^{2}\rightarrow m^{2} + 2 \epsilon/3)$.
By looking at the potential \eqref{eq:strangepotential} we see that the most stable vortex is identified by $|\phi_{2}|=1$, which in our notations is just the $\mathbb{C}P^{1}_{+}$ vortex.
In a typical situation expected to be realized in the interior of a neutron star, this potential is largely dominant as compared as the one generated by electromagnetic interactions, and the $\mathbb{C}P^{1}_{+}$ vortex turns out to be the most stable solution into which the other vortices must decay. Thus we expect that this kind of vortex plays a fundamental role for the physics of neutron stars.
%
%
%
\section{Boojums at the Interface}
The inner structure of neutron stars is still not completely clarified. However, it is now widely accepted that the outer region of the star core is in the $npe$ phase, where protons are superconducting, while neutrons are superfluid. The inner region may be characterized by the presence of hyperons or of the CFL phase. Here we consider the second possibility.

The  fast rotation of neutron stars is responsible for the formation of 
a triangular lattice of superfluid neutron vortices in the $npe$ phase and of color vortices in the CFL phase.
The internal structure of magnetic fields is not known well yet, but it is reasonable to expect that the  magnetic fields present at the surface of the star penetrate into the inner shells, where they are eventually confined into proton vortices when they reach the superconducting $npe$ phase.

\begin{figure}[ht]
\subfigure[]{
	\includegraphics[width=8cm,keepaspectratio=true]{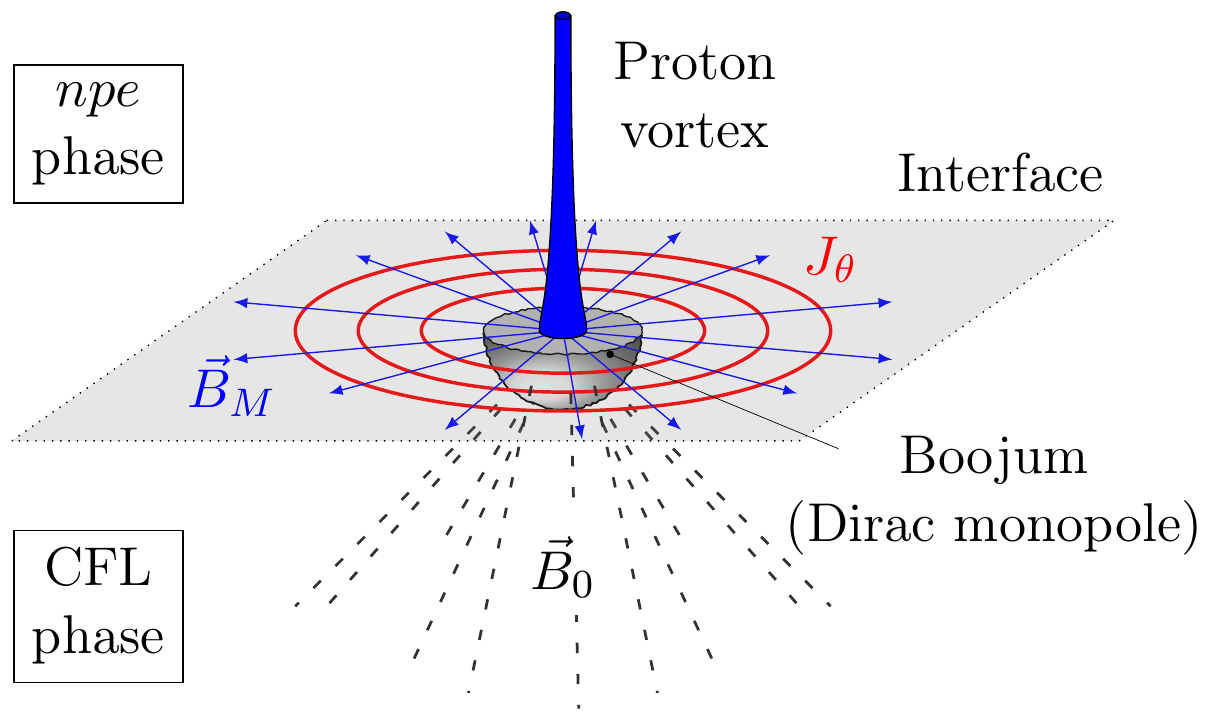}
		\label{BoojA}
}
\subfigure[]{ 
	\includegraphics[width=7cm]{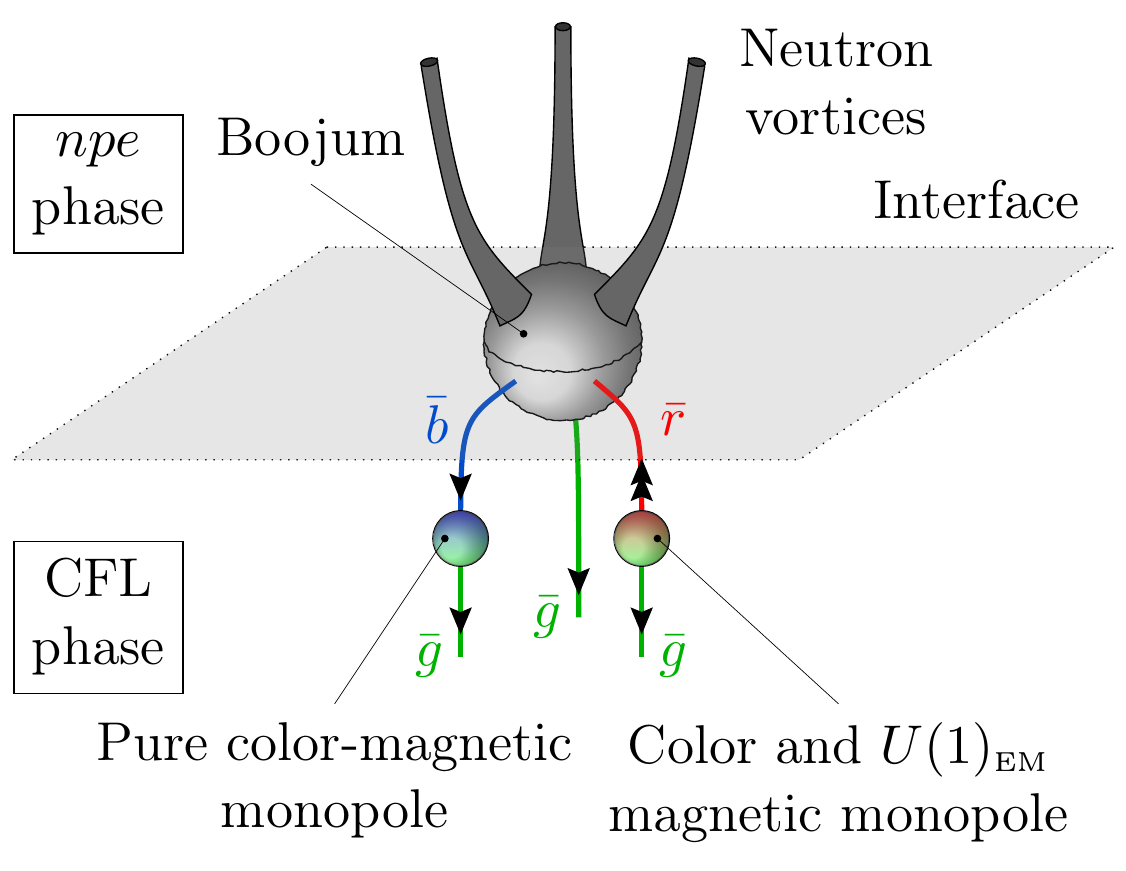}
		\label{BoojB}
}
\caption{\subref{BoojA} A superconducting proton vortex ending on the interface between the $npe$ and CFL phases.  A boojum forms at the contact point in the CFL phase region. The pure magnetic flux of the vortex splits into a $\vec{B}_{M}$ component, which is screened by a surface current and bent along the interface, and a $\vec{B}_{0}$ component emanating from the boojum, which looks like a Dirac monopole.
\subref{BoojB} Three neutron vortices ending in a boojum at the interface. The three BDM ($\bar{r}$), $\mathbb{C}P^{1}_{+}$ ($\bar{g}$) and $\mathbb{C}P^{1}_{+}$ ($\bar{b}$) vortices are depicted. The black arrows along the three vortices represent $U(1)_{\elm}$ magnetic flux. 
{The two monopole junctions described in the text are also depicted.}
}
\end{figure}

We now analyze the possible structures at the interface between the $npe$ and CFL phases, that can be created when  neutron and proton vortices hit the separation surface. 
Superconducting proton vortices carry a unit of magnetic flux $\Phi_{0}=\pi / e$. Moreover, they do not carry any unit of circulation. A proton vortex thus corresponds to a topologically trivial configuration once it enters the CFL phase. The proton vortex thus disappears  at the interface and a boojum is formed at the vortex ending point  in the CFL phase region. The disappearance of the proton vortex can be described as follows. When penetrating into the CFL phase, the $U(1)_{\elm}$ magnetic flux is converted into both the fluxes corresponding to the  massive and massless combinations $A_{M}$ and $A_{0}$. This is due to conservation of flux and to the the mixing (\ref{eq:mixing}) respectively. The massive combination $A_{M}$ is screened by a surface color-magnetic current circulating around the contact point. 
Unlike metallic superconductors, this is completely screened and cannot enter the CFL phase 
even if the flux is larger than the quantized flux of the non-Abelian vortex, 
because the non-Abelian vortex also has to carry the 1/3 quantized circulation. 
A rough estimate of the behavior of the current in proximity of the vortex endpoint can be obtained by using the London equation valid for an ordinary superconductor. Then we obtain 
$J_{\theta} \simeq \Phi_{0}/(2\pi r)$, where $\theta$ and $r$ are the planar polar coordinates centered at the contact point. On the other hand, the massless combination $A_{0}$ can spread freely into the CFL phase, being no superconducting currents that can screen it. 
This object looks like a Dirac monopole as is common in 
boojums in other systems such as helium superfluids.
This boojum  is qualitatively depicted in Fig. \ref{BoojA}.

Neutron vortices have a different physics at the interface compared to proton ones. A neutron vortex, being superfluid, carries a unit of quantized circulation. This implies that neutron vortices correspond to topologically non-trivial configurations in the CFL phase. In particular, three neutron vortices has to be attached to three semi-superfluid vortices, each of which carries 1/3 of the total circulation.
Since neutron vortices do not carry any $U(1)_{\elm}$ magnetic flux, color vortices in the junction have to come in triplet with vanishing  total  magnetic flux. The only possibility is the formation of a BDM, a $\mathbb{C}P^{1}_{+}$ and  a $\mathbb{C}P^{1}_{-}$ vortex. In fact, as explained in \cite{Vinci:2012mc}, the BDM vortex carries a $U(1)_{\elm}$ magnetic flux  
\begin{equation}
	\Phi^{\textsc{em}}_{\textsc{bdm}}=	\frac{\delta^{2}}{1+\delta^{2}}\frac{2 \pi}{e}\, , \quad \delta^{2} \equiv \frac23 \frac{e^{2}}{g_{s}^{2}} \, ,
\end{equation}
while the $\mathbb{C}P^{1}_{\pm}$ vortex flux is $\Phi^{\textsc{em}}_{\mathbb{C}P^{1}_\pm} = - \frac12 \Phi_{\textsc{em}}^{\textsc{bdm}}$. Moreover, the color neutrality imposes the formation of all the three kinds of vortices.  
When quarks travel around a vortex, they acquire a phase in general. Such phases 
have to match across the interface. The relevant order parameter in the $npe$ phase for neutron condensation is $\langle nn \rangle \sim \langle (qqq)(qqq) \rangle$, behaving like $\langle nn\rangle \sim e^{i \theta}$ in the presence of a neutron vortex of unit $U(1)_{\rm B}$ winding. From this,  we see that quarks get a phase $2\pi/6$, 
corresponding to $1/6$ $U(1)_B$ winding,
when they encircle a neutron vortex in the $npe$ phase.
In the CFL phase, the order parameter is $\langle qq \rangle$, which behaves $\langle qq \rangle \sim e^{i\theta}$ for a $U(1)_{\rm B}$ vortex or a triplet of semi-superfluid vortices, indicating that the quark fields 
get a phase $2\pi/2$ corresponding to $1/2$ $U(1)_{\rm B}$ winding, 
when they travel around a $U(1)_{\rm B}$ vortex 
or a triplet of  semi-superfluid vortices. 
Then, one concludes that three neutron vortices have to join at the boojum.

At a typical distance $\xi$ from the interface, though, the BDM and the $\mathbb{C}P^{1}_{-}$ solutions will ``decay'' to the $\mathbb{C}P^{1}_{+}$ vortex, due to their instability.  
We can estimate the length scale $\xi$ by referring to the low-energy effective action \eqref{eq:effaction}. 
Following
the same steps of \cite{Eto:2009tr} we obtain:
\begin{equation}
	\xi \sim m_{s}^{-1} \left( \frac\mu{\Delta_{\epsilon}} \right)^{2} \log\left( \frac\mu{\Delta_{\epsilon}} \right)^{-1/2}  \sim 4 {\rm GeV}^{-1} \, ,
\end{equation}
with 
the physical quantities being:
	$\mu \sim 10 {\rm GeV}, \Delta_{\epsilon} \sim 100 {\rm MeV}, K_{3} = 9$.
This length has to be compared with the thickness $d$ of the interface, which can be seen as a domain wall between the two different phases \cite{Giannakis:2003am}. Using the same values for physical parameters, we get $d \simeq 10^{-10} \xi$. Then the vortices decay at large distances from the interface. Since vortices decay into others with different fluxes, each junction corresponds in fact to a monopole. 
Unlike the Dirac monopole at the endpoint of a proton vortex, 
this is a confined monopole attached by vortices from its both sides. 
The monopole connecting $\mathbb{C}P^{1}_{-}$ to $\mathbb{C}P^{1}_{+}$ vortices is a pure color magnetic monopole, because the two vortices have the same $U(1)_{\elm}$ magnetic flux but different color-magnetic fluxes; the junction between BDM and $\mathbb{C}P^{1}_{+}$ is instead realized by a color-magnetic and $U(1)_{\elm}$ magnetic monopole, because for these vortices both the $U(1)_{\elm}$ magnetic and color-magnetic fluxes are different. The colorful boojum originated by neutron vortices is qualitatively depicted in Fig. \ref{BoojB}.

In order to present a more realistic structure of the neutron boojum in the CFL phase, we have  modeled the system as a regular lattices of boojums where the relative separation between vortices in a single boojum is denoted by $y(z)$ as a function of the distance $z$ from the interface. The center of the $i$-th boojum at the interface is indicated by $\vec x_{i}$. We use the Nambu-Goto action and approximate the interaction of vortices with that of global parallel vortices. The energy, per unit length, of the lattice is then:
\begin{align}
E_{\rm tot}(y)& =  3 \mathcal N \mathcal  T\sqrt{1+(dy/dz)^{2}}+V_{\rm int}(y),  \nonumber \\
V_{\rm int}(y)& =  -3 \mathcal N \mathcal  T  \log |y|-9 \mathcal  T \sum_{i>j}\log |\vec x_{i}-\vec x_{j}| \, , 
\end{align}
where $\mathcal N$ is the number of boojums and $\mathcal T$ is the tension of a color-magnetic vortex. The first term in the potential above is the interaction energy between vortices in the same boojum while the second term represents the one between boojums, where the shape of the boojum has been neglected. Notice that the first term in the potential has to be regularized in the limit $y\rightarrow 0$. 
\begin{figure}[ht]
	\subfigure[]{
   \includegraphics[width=10cm]{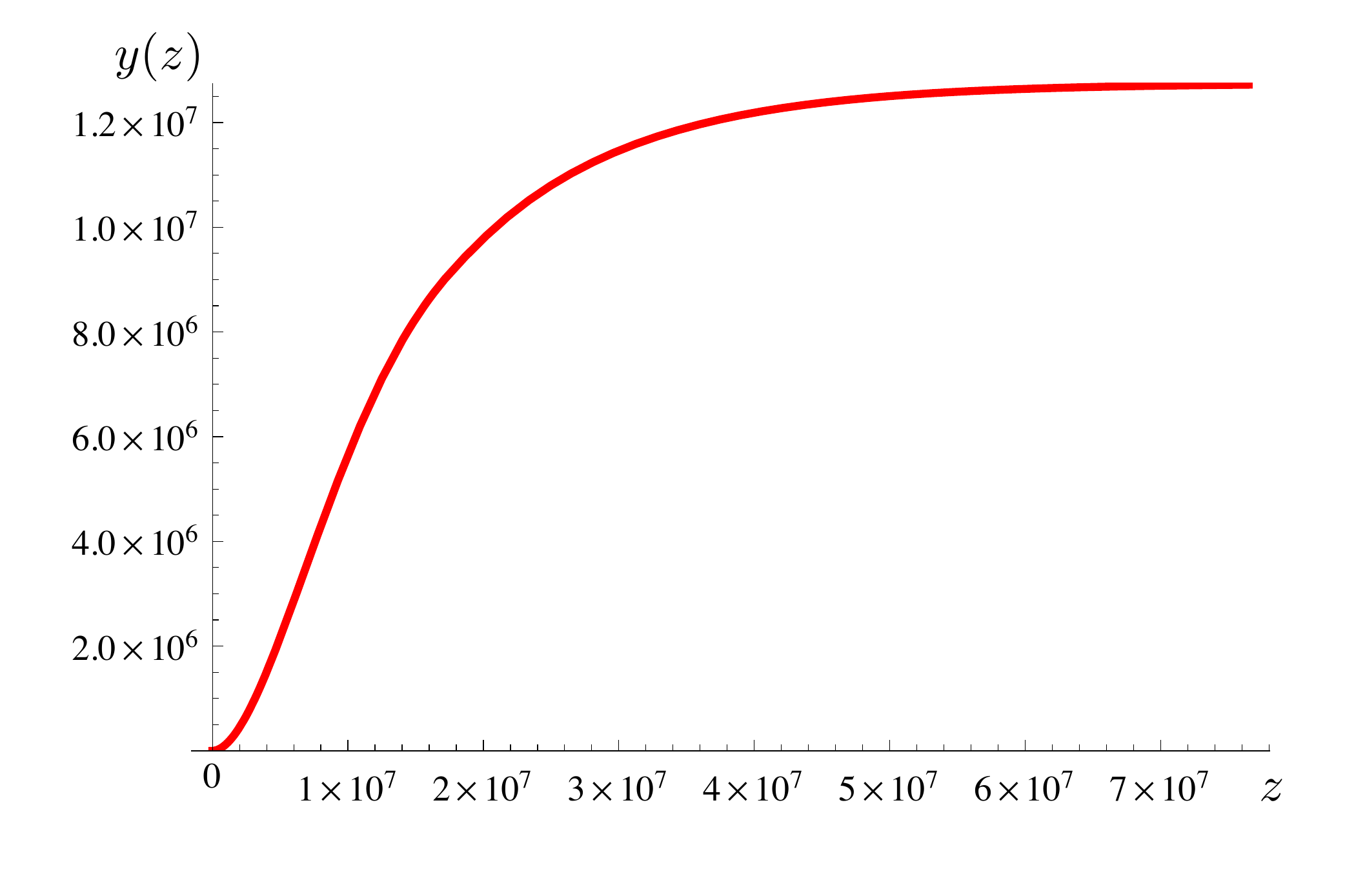}
		\label{fig:2DBoojum}
}
    \subfigure[]{
    \includegraphics[width=10cm]{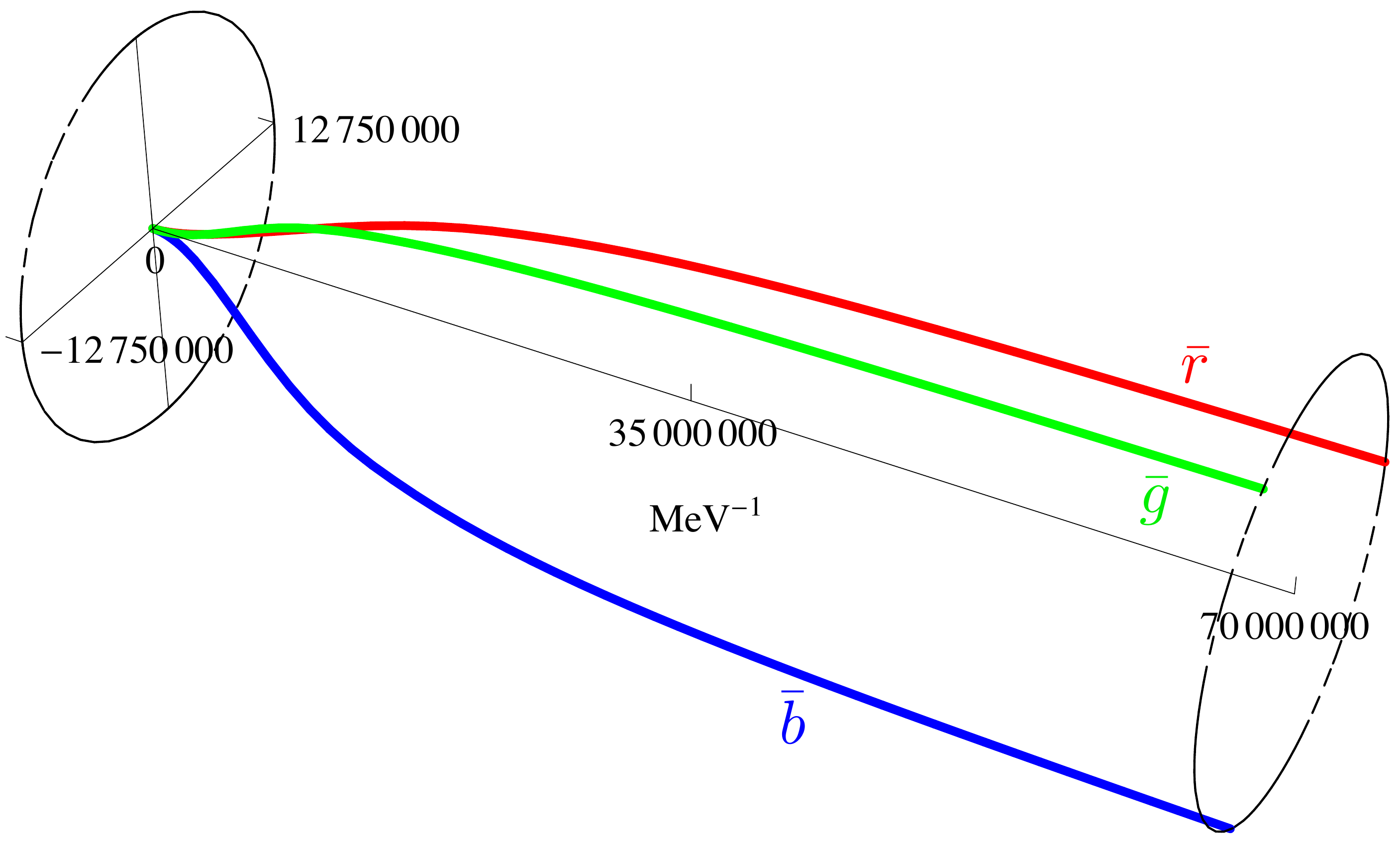}
	\vspace{-0.8cm} 
	\label{fig:3DBoojum}
}
	\caption{\subref{fig:2DBoojum} Transverse shape and \subref{fig:3DBoojum} 
three dimensional shape of a neutron boojum. The $npe$-CFL interface is on the left, where three color vortices are originated by three neutron vortices. We have used typical values of the GL parameters as done, for example, in Ref.~\cite{Vinci:2012mc}. 
In \subref{fig:3DBoojum}, we have ignored the effect of the strange quark mass 
for illustration purpose so that the vortices remain colored. 
}
\label{fig:Boojum}
\end{figure}
The numerical shape of the boojum is shown in Fig.~\ref{fig:Boojum},  where we have evaluated the interaction potential for a one dimensional lattice for simplicity. The most important property of the boojum that can be inferred from this very simplified numerical analysis is that the longitudinal  size scales proportionally to the transverse lattice spacing. We have also checked that this property and the shape of the boojum also do not depend on the choice of the regularization of the interaction potential for coincident vortices.

\section{Conclusions}

In summary we have found that 
when neutron superfluid and proton superconducting vortices 
hit the interface between the $npe$ and CFL phases, 
two different boojums appear at these ending points.
The magnetic flux of  a proton vortex is decomposed into fluxes of
the massive and massless combinations of gauge fields in the CFL phase. 
The former is  screened and expelled completely from the CFL phase by formation of color-magnetic surface currents, while the latter can spread freely into the CFL phase 
with the boojum as a source (a Dirac monopole).
A different boojum conects three neutron vortices to three non-Abelian vortices carrying different color magnetic flux with the total flux canceled out at the contact point. Two of these vortices turn into the stable one through confined monopole junctions. One of these confined monopoles has a purely color magnetic charge; the other has also $U(1)_{\elm}$ magnetic charge. 
We finally have modeled the system to derive the shape of the boojum in the CFL phase.

The system we described is the simplest configuration in the pure CFL phase, which is realized at very high densities $\mu \gg m_s^2/\Delta_{\textsc{cfl}}$. However, in a more realistic setting, the strange quark mass stresses the CFL phase and other phases have to be considered. For example, when the chemical potential is low enough that $m_s\gtrsim m_{u,d}^{1/3}\Delta_{\textsc{cfl}}^{2/3}$, kaons  can condense, leading to the so called  CFL-K$^0$ phase \cite{Shafer:2000}, where other kinds of vortices arise \cite{Kaplan}.  However in the high density regime the boojum can be modified only in the vicinity of the interface, while the overall structure of the junction is kept unchanged.
We leave the complication of considering kaon condensation and other phases for future works.

W.V. thanks the FTPI Institute of the University of Minnesota for the nice hospitality while part of this work was completed. 
M.N. thanks M.~Eto for a discussion at the early stage and 
INFN, Pisa, for partial support and hospitality while this work was done.
The work of M.N. is supported in part by 
Grant-in Aid for Scientific Research (No.~23740198) 
and by the ``Topological Quantum Phenomena'' 
Grant-in Aid for Scientific Research 
on Innovative Areas (No.~23103515)  
from the Ministry of Education, Culture, Sports, Science and Technology 
(MEXT) of Japan. 


\end{document}